\documentclass[twocolumn,prl,aps,superscriptaddress,showpacs,amsmath,amssymb,floatfix,longbibliography]{revtex4-1}

\usepackage{graphicx}
\usepackage{amsmath}
\usepackage{siunitx}
\usepackage{dcolumn}
\usepackage{bm}

\usepackage{bbm}
\usepackage{float}
\usepackage[colorlinks,
linkcolor=red,
anchorcolor=blue,
urlcolor=blue,
citecolor=green]{hyperref}

\begin{document}


\title{Nonadiabatic  dynamics and geometric phase of an ultrafast rotating electron spin} 


\author{Xing-Yan Chen}
\affiliation{Center for Quantum Information, Institute for Interdisciplinary Information Sciences, Tsinghua University, Beijing 100084, China}
\affiliation{Max-Planck-Institut f\"ur Quantenoptik, Garching 85748, Germany}
\affiliation{Fakult\"at f\"ur Physik, Ludwig-Maximilians-Universit\"at M\"unchen, M\"unchen 80799, Germany}

\author{Tongcang Li}
\affiliation{Department of Physics and Astronomy, Purdue University, West Lafayette, Indiana 47907, USA}
\affiliation{School of Electrical and Computer Engineering, Purdue University, West Lafayette, Indiana 47907, USA}
\affiliation{Purdue Quantum Center, Purdue University, West Lafayette, Indiana 47907, USA}
\affiliation{Birck Nanotechnology Center, Purdue University, West Lafayette, Indiana 47907, USA}
    
\author{Zhang-qi Yin}\email{yinzhangqi@mail.tsinghua.edu.cn}
\affiliation{Center for Quantum Information, Institute for Interdisciplinary Information Sciences, Tsinghua University, Beijing 100084, China}


\date{\today}

\begin{abstract}
The spin in a rotating frame has attracted a lot of attentions recently, as it deeply relates to both fundamental physics such as  pseudo-magnetic field and geometric phase, and applications such as gyroscopic sensors. However, previous studies only focused on  adiabatic limit, where the rotating frequency  is much smaller than the spin frequency. Here we propose to use a levitated nano-diamond with a built-in nitrogen-vacancy (NV) center to study the dynamics and the geometric phase of a rotating electron spin without adiabatic approximation. We find that the transition between the spin levels appears when the rotating frequency is comparable to the spin frequency at zero magnetic field. Then we use Floquet theory to numerically solve the spin energy spectrum, study the spin dynamics and calculate the geometric phase under a finite magnetic field, where the rotating frequency to fulfill the resonant transition condition could be greatly reduced. 

\end{abstract}


\maketitle 


The electron and nuclear spins in a rotating frame deeply connect to both fundamental physics and applications.
The frequencies of the spins will shift in the rotating frame, which can be explained by an emerged pseudo-magnetic field \cite{Barnett1915,Barnett1935}.
The quantum mechanical geometric phase was also predicted to appear in these systems in adiabatic limit, where the frequency of rotating frame is much less than the frequencies of the spins   \cite{Ajoy2012,MacLaurin2012,Ledbetter2012}. The pseudo-magnetic field has been detected by both  nuclear \cite{Chudo2014} and  electron spins \cite{Wood2017,Wood2018}.
However, the geometric phase which is proportional to the rotating frequency and can be used as a gyroscopic sensor \cite{zhang2016inertial}, has been too small to be measured in a traditional mechanical rotor with a maximum rotation frequency of about $10$ kHz \cite{Wood2018}. 

Here we propose to use a levitated nanodiamond that can be driven to rotate at an ultrahigh speed in vacuum to study the geometric phase of a rotating electron spin. Our proposal is based on recent breakthroughs in levitated optomechanics \cite{Yin2013b,Chang2010,Romero2010,Yin2013,Scala2013b,Shi2016,Hoang2016a,Ma2017,Arita2013,Kuhn2017,Monteiro2018,PhysRevLett.121.040401,Zhao2014,Ranjit2016,Monteiro2017}. Nanodiamonds with nitrogen-vacancy centers that host electron spins have been levitated in vacuum with optical tweezers \cite{Neukirch2015a,Hoang2016}, ion traps\cite{doi:10.1021/acs.nanolett.8b01414,PhysRevLett.121.053602}, and magneto-gravitational traps \cite{hsu2016cooling}. Recently, rotation frequencies larger than $1$ GHz have been experimentally observed with optically levitated nanoparticles driven by circularly-polarized lasers \cite{Ahn2018,Reimann2018}. In this way, for the first time, the frequency of a mechanical rotor approaches  the frequency of the electron spin in the NV center. This will generate a large geometric phase. Previous studies based on adiabatic approximation will no longer be valid  \cite{Ajoy2012,MacLaurin2012,Ledbetter2012}. 
A theory of nonadiabatic spin dynamics and geometric phase in a rotating frame is needed.

In this paper, we study the  electron spin dynamics and calculate the quantum geometric phase of an NV center in an ultra-fast rotating levitated nanodiamond without adiabatic approximation. 
We find that transitions between the spin energy levels appear when the angle $\theta$ between the axis of the NV center and the axis of the rotor is not zero. This effect is negligible in adiabatic limit, but becomes important in the nonadiabtic regime. 
By clockwise (counterclockwise) rotating the nano-diamond, the resonant Rabi oscillation between $|0\rangle$ and  $|+1\rangle$ ($|-1\rangle$) of the NV center could realize, if the rotational frequency approaches the frequency of electron spin in the NV center without external magnetic field. We calculate 
the quantum geometric phase of the electron spin, which is in consistent with the previous studies in adiabatic limit  \cite{Ajoy2012,MacLaurin2012,Ledbetter2012}, and is maximized near the resonant point.  The energy spectrum, dynamics, and nonadiabatic geometric phase of the electron spin under the finite magnetic field are numerically solved. 
We find that the resonant transition could be achieved with much lower rotating frequency with an external magnetic field.

\begin{figure}
	\centering
	\includegraphics[width=0.8\linewidth]{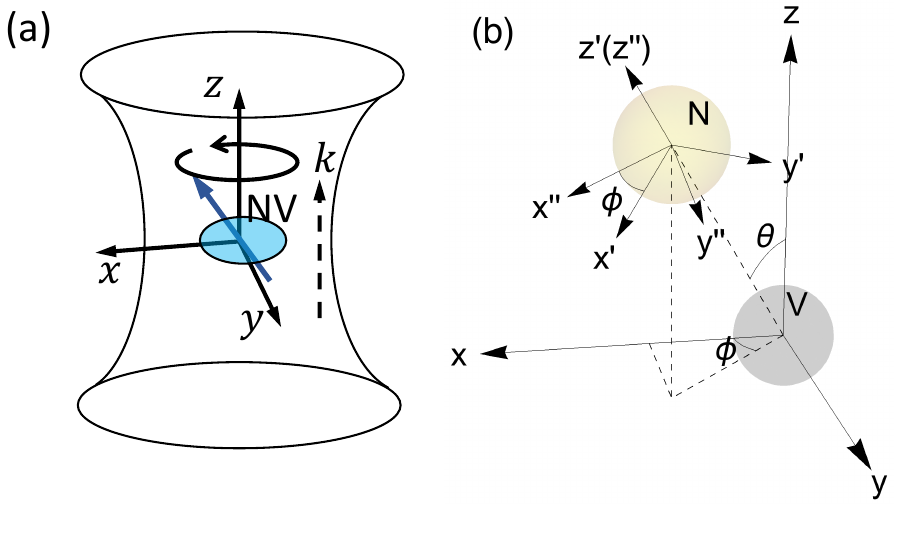}
	\caption{(a) A nanodiamond with a build-in  NV center is levitated in an optical trap. A circular polarized laser drives the nanodiamond to rotate at angular frequency $\omega$.
	(b) The frames $Ox'y'z'$ and $Ox''y''z''$ are defined by the rotational transformations $R(t)=R_z(\phi(t))R_y(\theta)$ and $W(t)= R(t)R_z(-\phi(t))$, respectively. 
The rotating spin states are defined to be static in the frame $Ox''y''z''$.
}
	\label{fig:spin}
\end{figure}

We consider a non-spherical nano-diamond optically trapped in high vacuum. The length of the three axes of the nanodiamond are different. Therefore, its rotational degrees of freedom could be manipulated by the driving laser.
We adopt the polarization of the driving laser to be circular. The nano-diamond could be driven to rotate at a constant angular velocity $\bm{\omega}$ \cite{Ahn2018,Reimann2018}. There is a nitrogen-vacancy center, with electron spin $S = 1$, in the nanodiamond. 
As shown in Fig. \ref{fig:spin}(a), we choose the direction of $\bm{\omega}$ along z-axis, and define spherical coordinates $\theta$  and $\phi(t) = \omega t$. We denote $\theta$ as the angle between the rotational axis and the axis of the NV center. 
For simplicity, we consider first the dynamics of the electron spin without external magnetic field. The Hamiltonian of the rotating NV center can be obtained by conducting a rotational transformation $R(t) = R_z(\phi(t))R_y(\theta)$ on the stationary Hamiltonian $H_0 = D S_z^2$ ($\hbar =1$), which reads 
	$H(t) = R(t)H_0 R^\dagger(t)$.
Here the rotation of spin-1 by the angle $\alpha $ along direction $\bm{n}$ is given by $R_{\bm{n}}(\alpha) = e^{-i\alpha \bm{n}\cdot\bm{S}}$.  

The Hamiltonian $H(t)$ is written in the $S=1$ basis in the inertial (lab) frame. Alternatively, we can rewrite the Hamiltonian in the basis which rotate with the solid spin and study the dynamics of a rotating spin. The unitary transformation from the static basis to the rotating basis is given by $W(t) = R(t)R_z(-\phi(t))$, which differs from $R(t)$ by an additional $R_z(-\phi(t))$ rotation. The rotational transformation corresponding to $R(t)$ and $W(t)$ are shown in Fig. \ref{fig:spin}(b), which are denoted by $Ox'y'z'$ and $Ox''y''z''$, respectively. The additional term $R_z(-\phi(t))$, which cancels the rotation of the local orthogonal coordinates, moves the geometric phase into the dynamical phase \cite{Giavarini1989,SM}. To see that, we write down the Hamiltonian after the unitary transform $\tilde{H}(t) = W(t)H(t)W^\dagger(t) + i\hbar \partial_t W(t)W^\dagger(t)$ as
\begin{equation} \label{eq:HT}
    \tilde{H}(t)= H_0 + \omega(1 - \cos \theta)S_z
	 - \frac{\omega}{2} \sin\theta [e^{-i(\omega t + \phi_0)}S_+ + h.c.],
\end{equation} 
where $S_\pm = S_x \pm i S_y$. The constant phase $\phi_0$ in Eq. \eqref{eq:HT} arises as $Ox'y'z'$ and $Ox''y''z''$ in Fig. \ref{fig:spin}(b) have a relative rotation $R_z(-\phi_0)$. 

In the interaction picture given by the unitary transformation $U = e^{i\omega S_z t}$, the time-independent Hamiltonian reads
\begin{equation}\label{eq:H_I}
	\tilde{H}_I =
	\begin{pmatrix}
	D - \omega\cos \theta & -\Omega^*/2 & 0 \\
	-\Omega/2 & 0 & -\Omega^*/2 \\
	0 & -\Omega/2 & D + \omega \cos \theta
	\end{pmatrix},
\end{equation}
where we denote $\Omega = \sqrt{2}\omega e^{i\phi_0}\sin\theta$ as the Rabi frequency. In the rest of this paper, the phase $\phi_0$ of the Rabi frequency is eliminated by redefining the $S_z$ states.  

Let us solve the Hamiltonian \eqref{eq:H_I} in two limit at first,  the adiabatic limit $\omega \ll D$ and the near resonant limit $|D \pm \omega \cos \theta| \ll \Omega$. 
In adiabatic limit, the effect of transitions between spin states is negligible. 
We can neglect the off-diagonal terms in Hamiltonian  \eqref{eq:HT} and get the effective Hamiltonian $\tilde{H}_e = D S_z^2 + \omega(1 - \cos \theta)S_z$,
where the last term $\omega(1 - \cos \theta) S_z$ is called the rotating induced level shift (RILS) term. It is  consistent with the previous studies on  the adiabatic geometric phase \cite{Ajoy2012,MacLaurin2012,Ledbetter2012}. When  $|D \pm \omega \cos \theta| < \Omega$, the off-diagonal terms in the Hamiltonian \eqref{eq:H_I} could induce transitions between  $|0\rangle$ and $|\pm 1\rangle$. The resonant condition requires the angular frequency $\omega_\pm = \pm D/\cos\theta$. At resonance, and in the  limit $\sin\theta \ll 1$, we can ignore off-resonant terms and get perfect Rabi oscillation in a two-dimensional subspace with the Rabi frequency $\Omega$ \cite{SM}. 
For rotating frequency  $\omega=\omega_+$,  the resonant transition between $|0\rangle$ and $|+1\rangle$ happens, while for $\omega=\omega_-$ the resonant transition between $|0\rangle$ and $|-1\rangle$ appears. This driving selectivity comes from the conservation of angular momentum.

\begin{figure}
	\centering
       \includegraphics[width=1.0\linewidth]{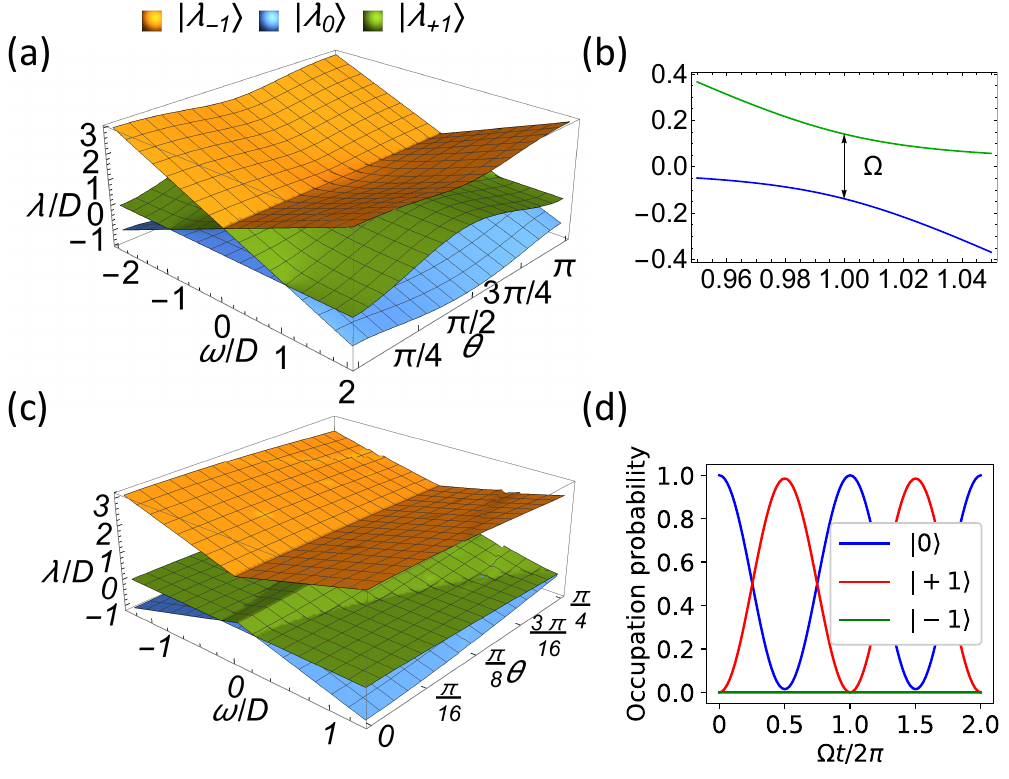}
	\caption{(a) The quasi-energies $\lambda$ as a function of rotating frequency $\omega$ and $\theta$. When near crossover $\omega = \pm D/\cos\theta$, the eigenstates of $S_z$ will be significantly mixed, which means that there will be Rabi oscillation. The legends means that the quasi-state starts with the respective eigenstates of $S_z$. (b) is the scaled slice with $\theta = \pi/100$ in (a). The quasi-energy splitting gives the Rabi frequency $\Omega$. (c) and (d) are the the numerical solution of quasi-energies and time-evolution in the presence of a static magnetic field with $\Delta = 0.803D$. The magnetic field satisfies the resonant condition with $\theta = \pi/100$ and $\omega = 0.2D$ in (d). The quasi-energies are obtained by diagonalizing the Floquet Hamiltonian.}
	\label{fig:quasi}
\end{figure}

For an arbitrary angle $\theta$, we need to diagonalize the whole $3\times3$ matrix of the Hamiltonian \eqref{eq:H_I}. The quasi-energies are given by the solution of the cubic equation
\begin{equation}
	\lambda^3 -2D \lambda^2 -(\omega^2 - D^2)\lambda + \omega^2D\sin^2\theta = 0,
\end{equation}
which are  $\lambda_0$ and $\lambda_{\pm 1}$.  
We denote the Floquet states with quasi-energies $\lambda_n$ by $|\lambda_n\rangle$, with $n = 0,\pm1$. Here,  $|\lambda_n\rangle$ smoothly connects to $|n\rangle$ at the adiabatic limit $\omega \ll D$.
The quasi-energy spectrum as a function of $\omega$ and $\theta$ is shown in Fig. \ref{fig:quasi}(a).
The quasi-energy spectrum $\lambda$ has a level crossing around $\omega = \pm D$ if $\theta=0$. As long as $\theta \neq 0$, the quasi-energy spectrum has an 
avoided level crossing near the resonant point  $\omega_\pm$, as shown in Fig. \ref{fig:quasi}(b). The quasi-energy splitting at the resonant point gives the Rabi frequency $\Omega$. In Fie. \ref{fig:quasi}(a), there is also an avoided level crossing between $|+1\rangle$ and $|-1\rangle$ near $\theta = \pi/2$, which corresponds to a second order effective Rabi oscillation between these two states with Rabi frequency $\omega^2/D$ \cite{SM}.
We plot the dynamics of the electron spin in a NV center in rotating frame in Ref. \cite{SM}.

If the NV center is rotating in the presence of an non-zero external magnetic field, the total Hamiltonian reads 
$H(t) = R(t)(DS_z^2 + g \mu_B B \bm{S}\cdot\bm{n})R^\dagger(t) 
 = R(t)H_1 R^\dagger(t)$ \cite{SM},
where $\bm{n} = (\sin\theta,\cos\theta,0)$ is the unitary vector along the magnetic field direction in the rotating frame, and 
$\Delta = -g\mu_B B$. Similar to Eq. \eqref{eq:HT}, we apply the unitary transformation $\tilde{H}_1(t) = W(t)H_1(t)W^\dagger(t) + i\hbar \partial_t W(t)W^\dagger(t)$ and get
\begin{equation} \label{eq:HTB}
	\begin{split}
	\tilde{H}(t)	& = DS_z^2 - \Delta\cos\theta S_z + \Delta\sin\theta S_x  \\ &+ \omega(1 - \cos \theta)S_z 
	 - \frac{\omega}{2} \sin\theta(e^{-i\omega t}S_+ + h.c.).
	\end{split}
\end{equation}

The presence of the magnetic field change the eigenstates of the spin, which are no longer the eigenstates of operator $S_{z}$. When the spin is rotating, the time-independent eigenstates will be further changed.
 Therefore, it is quite difficult to analytically solve the problem.  However, in the limit of $\theta \ll 1$, the misalignment of the magnetic field to the spin is negligible and the eigenstates are nearly not changed. Therefore, for simplicity, we consider $ 0 < \Delta  < D$ and take the small angle limits $\theta \ll 1 - \Delta/D$, and only consider the nearly resonant situation. 
 The effective Hamiltonian (to the order of $\theta^2$) reads  
\begin{equation} \label{eq:HTBs}
	\tilde{H}(t)  = \tilde{D}S_z^2 -\tilde{\Delta}S_z 
	- \frac{\omega}{2}\theta(e^{-i\omega t}S_+ + h.c.),
\end{equation}
where $\tilde{D} = D + \frac{3D\Delta^2}{2(D^2-\Delta^2)}\theta^2$ and $\tilde{\Delta} = \Delta - \frac{1}{2}\omega \theta^2 - \frac{D^2\Delta}{2(D^2-\Delta^2)}\theta^2 $.
The resonant condition is given by $\tilde{D} \mp \tilde{\Delta} = \pm\omega$. 
Therefore, the magnetic field compensate to $\omega$ and allows us to observe Rabi oscillation at a lower angular frequency. 

When the angle $\theta$ is not approaching zero, the above perturbative analysis becomes invalid,  and we adopt the Floquet formalism \cite{Grifoni1998,Shirley1965} to numerically solve the evolution of Eq. \eqref{eq:HTB} \cite{SM}. Since the quasi-energies are determined uniquely only up to a multiple of $\hbar\omega$ \cite{SM}, the branches start at the same quasi-energy with slopes $+ n \hbar\omega$ and represent the same Floquet states with quasi-energy $+ n \hbar\omega$. For example, an avoided level crossing between $|\lambda_0\rangle$ start with slope $0$ and $|\lambda_{+1}\rangle$ start with slope $-\hbar\omega$ means there is strong transition from $|0\rangle$ to $|+1\rangle$ by absorbing one photon. In order to compare with quasi-energies under the zero magnetic field, as shown in Fig. \ref{fig:quasi}(a), we choose the three quasi-energy branches that smoothly connected to the $\omega = 0$ eigenvalues with slopes equal to $\omega$, $0$, and $-\omega$ for $|\lambda_{-1}\rangle$, $|\lambda_0\rangle$, and $|\lambda_{+1}\rangle$, respectively.

We numerically solve the quasi-energies with external magnetic. 
As shown in Fig. \ref{fig:quasi}(c), there is an avoided level crossing between $|\lambda_{0}\rangle$ and $|\lambda_{+1}\rangle$ for $\theta\neq 0$. 
The resonant angular frequency $\omega$ increases if the angle between the spin and the rotating axis $\theta$ increases. 
The quasi-energy splitting corresponds to the Rabi frequency $\Omega = \sqrt{2}\omega\sin\theta$. If there is no magnetic field, for $\theta = \pi/100$ and $\omega = 0.2~ D$, the Rabi oscillation between $|0\rangle$ and $|+1\rangle$ is negligible. By applying a static magnetic field with $\Delta = 0.803~D$ to meet the resonant condition, as shown in Fig.\ref{fig:quasi}(d),  there is an almost perfect resonant Rabi oscillation between $|0\rangle$ and $|+1\rangle$. In this way, the resonant electron spin dynamics could be realized with rotating frame frequency $\omega$ much lower than $D$.

\begin{figure}
	\centering
	\includegraphics[width=0.75\linewidth]{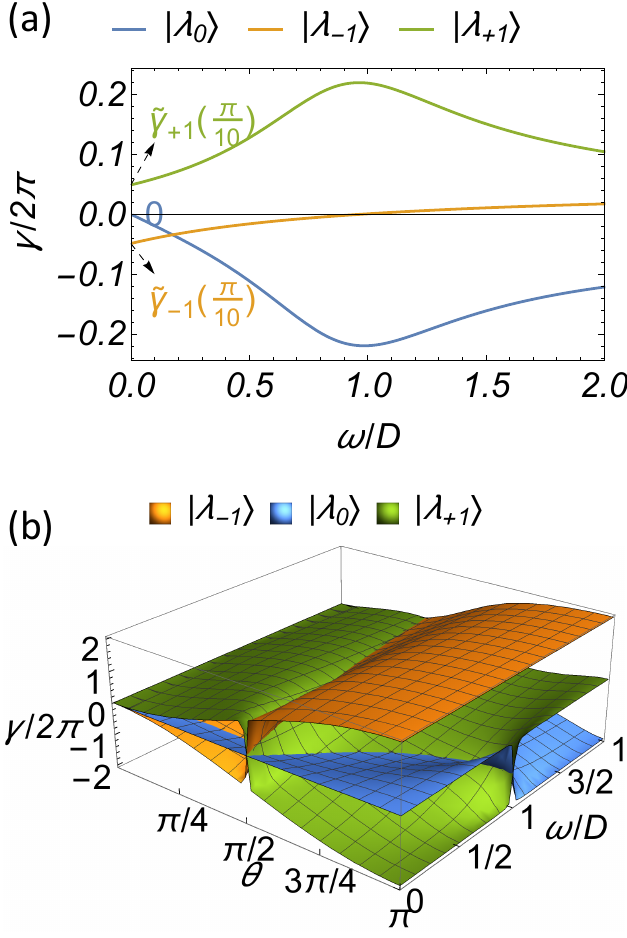}
	\caption{(a) Non-adiabatic geometric phases $\gamma_{0,\pm}$ for the three cyclic states $\lambda_{0,\pm}$ under the angle $\theta = \pi/10$ and  different rotating frequency $\omega$.  At adiabatic limit $\omega \ll D$, the geometric phases are given by $\tilde{\gamma}_{\pm 1}(\theta) = \pm 2\pi(1-\cos\theta)$. At resonance $\omega = D/\cos\theta$,  the geometric phases are $ \gamma_{+1}, ~\gamma_{0},~ \gamma_{-1} = \sqrt{2}\pi\sin\theta,~ -\sqrt{2}\pi\sin\theta,~ 0$. 
(b) Non-adiabatic geometric phases $\gamma$ for the three cyclic states under different rotating frequency $\omega$ and angle $\theta$. }
	\label{fig:gp}
\end{figure}

Based on the Floquet formalism \cite{SM},
we can derive the non-adiabatic geometric phase for the electron spin of the NV center in a ultra-fast rotating nanodiamond. 
 The non-adiabatic geometric phases for the cyclic states are defined as \cite{PhysRevLett.58.1593,Moore1990,Moore1990a}, 
\begin{equation}\label{eq:gp}
\gamma_n = i \int^T_0 \langle\lambda_n|W(t)\frac{d}{dt}W^\dagger(t)|\lambda_n \rangle dt,
\end{equation}
where $W(t)$ is chosen in order to recover the RILS dynamical phase shift back into the geometric phase. Let us consider the case without external magnetic field first. The Eq. \eqref{eq:gp} can be rewrote as 
$\gamma_n =   \int^T_0 \langle\lambda_n|\tilde{H}_I - DS_z^2 + \omega S_z |\lambda_n\rangle dt$, and plug in the coefficient of Floquet states, we get
\begin{equation}\label{eq:gpc}
	\gamma_n = \frac{2\pi}{\omega}(\lambda_n - (D-\omega)|c_{n,+1}|^2 - (D+\omega)|c_{n,-1}|^2),	
\end{equation}
where $c_{n,k}$, $k = 0,\pm1$ are the coefficients of $|\lambda_n\rangle$ in the spin basis. Based on the simplified Hamiltonian in the limit $\omega \ll D$ and $\omega \sim D/\cos\theta$, we can obtain the geometric phase using Eq. \eqref{eq:gpc}. If the rotation is adiabatic  $\omega \ll D$, the Floquet states $|\lambda_n\rangle$ has almost no mixing between the spin states. The quasi-energies  $\lambda_{+1},~\lambda_{0},~\lambda_{-1} = D + \omega\cos\theta,~ 0, ~D - \omega\cos\theta$, with corresponding geometric phases $\gamma_{+1},~ \gamma_{0},~ \gamma_{-1} = 2\pi(1-\cos\theta),~ 0,~  -2\pi(1-\cos\theta)$, which are consistent with the previous studies \cite{Ajoy2012,MacLaurin2012,Ledbetter2012}.
 In the resonant regime with $ \omega \simeq D/\cos\theta$ and $\theta \ll 1$, there is strong mixing between spin states. The corresponding Floquet states are $|\lambda_{+1}\rangle = (|0\rangle + |+1\rangle)\sqrt{2}$, $|\lambda_{0}\rangle = (|0\rangle - |+1\rangle)\sqrt{2}$, and $|\lambda_{-1}\rangle =|-1\rangle$ with quasi-energies $ \lambda_{+1},\lambda_{0},\lambda_{-1} = \Omega/2,-\Omega/2,2D$. The non-adiabatic geometric phases are  $ \gamma_{+1}, ~\gamma_{0},~ \gamma_{-1} = \sqrt{2}\pi\sin\theta,~ -\sqrt{2}\pi\sin\theta,~ 0$.  As shown in Fig. \ref{fig:gp}(a), a slightly detune from resonance will reduce the geometric phase, which means that for small angles $\theta$ the geometric phase will maximize at resonance. For general situations with arbitrary $\omega$ and $\theta$, we provide numerical result in Fig. \ref{fig:gp}(b). The analysis for limit cases indicates that there is crossing between the two limits. Also, the peak behavior at resonance $\omega = D/\cos\theta$ is demonstrated. When $\theta$ increases, the geometric phases  become larger, but the peak is not so apparent due to the break down of two-level approximation at the large angle. From Fig. \ref{fig:gp}(b), the crossing at $\theta = \pi/2$ corresponds to the second order Rabi oscillation between $|\pm1\rangle$, and the crossing at $\theta \sim \pi$ and $\omega \simeq D$ corresponds to the Rabi oscillation similar to the small angle \cite{SM}. 

From Eq. \eqref{eq:gpc} we can also reveal the relation between quasi-energy $\lambda_n$ and geometric phase $\gamma_n$ for small $\theta$. In adiabatic limit, the geometric phase is identical to the phase of RILS term accumulating in a single period. At resonance regime, the geometric phases for the two resonant states are given by their Rabi frequency accumulate in a single period. In these two situations, the measurement of the geometric phase and the quasi-energy are equivalent. 

The adiabatic geometric can be used for measuring rotating frequency in the adiabatic limit \cite{Ajoy2012,MacLaurin2012,Ledbetter2012}. Our analysis shows that this method still works in the nonadiabatic regime $\omega \sim D$, where the nonadiabatic geometric phase can be measured with spectroscopic or interference \cite{Anandan1992,Appelt1995,DAS2005318}. 
Moreover, as the rotating frequency can be determined with high precision by measuring the scattering photon of the nano-diamond \cite{Ahn2018,Reimann2018}, the angle $\theta$ could be measured through the Floquet quasi-energy spectrum  \cite{PhysRevB.46.14675,PhysRevLett.105.257003,shu2018observation}. In the limit $\theta \ll 1$, the measurement of quasi-energies is equivalent to measure the Rabi frequency $\Omega$. The uncertainty of angular measurement $\delta\theta = \delta \Omega/(\sqrt{2}\omega\cos \theta)$ is inversely proportional to rotating frequency $\omega$ and minimized around $\theta =0$.

Under external magnetic field, the non-adiabatic geometric phase given by Eq. \eqref{eq:gp} is similar to the case without magnetic field \cite{SM}. For simplicity, we only analyze the limit case where $\omega \sim 0$ and near resonance, and for small angle $\theta$ where the Hamiltonian is given by Eq. \eqref{eq:HTBs}. From Eq. \eqref{eq:gp}, when apply to $\omega \sim 0$ case the geometric phases are $2\pi(1-\cos\theta)S_z$ which are the same as zero field. This is because the magnetic field only shift the energy level therefore only affects the dynamical phases. At resonance, the geometric phases for the two resonant states are proportional to the Rabi frequency, given by $\pm\sqrt{2}\pi\sin\theta$ which are also the same as zero field. 

We briefly discuss the experimental feasibility.  
As the silica-based nano-particles have been driven to the GHz rotating frequency regime, we believe that it is also possible to optically drive the nanodiamond to rotate in GHz. The main obstacle is the optical heating of the NV center in diamond, which could be resolved by adopting pure diamond \cite{Frangeskou2018} and using nano-refrigerator \cite{Rahman2017}. The Rabi frequency induced by GHz rotating nanodiamond is much larger than MHz. The dephasing time of the NV center in a nanodiamond is in the order of $\mu$s or longer \cite{Knowles2014}. Therefore, the rotating induced Rabi oscillation or the avoided level crossing could be observed.  

In conclusion, we study the electrons spin dynamics and geometric phase in a levitated ultra-fast rotating nanodiamond, without adiabatic approximation. The  Rabi oscillation appears if the rotating frequency matches the electron spin levels splitting, even without an external magnetic field. 
We define and calculate the nonadiabtic geometric phase of the eletron spin in a rotating frame, which could be used for an angular sensor.
We think that the similar phenomena may also appear in the nuclear spins in a rotating frame, with much lower rotating frequency. 

\begin{acknowledgments}
Z.Q.Y. is supported by National Natural Science Foun-
dation of China NO. 61771278, 61435007, and the Joint
Foundation of Ministry of Education of China (6141A02011604).
T.L. is supported by NSF under Grant No. PHY-1555035.  
We thank the helpful discussions with Nan Zhao and Ying Li.
\end{acknowledgments}

	
%

\end{document}



\title{SUPPLEMENTAL MATERIAL \\ Nonadiabatic  dynamics and geometric phase of an ultrafast rotating electron spin} 


\author{Xing-Yan Chen}
\affiliation{Center for Quantum Information, Institute for Interdisciplinary Information Sciences, Tsinghua University, Beijing 100084, China}
\affiliation{Max-Planck-Institut f\"ur Quantenoptik, Garching 85748, Germany}
\affiliation{Fakult\"at f\"ur Physik, Ludwig-Maximilians-Universit\"at M\"unchen, M\"unchen 80799, Germany}

\author{Tongcang Li}
\affiliation{Department of Physics and Astronomy, Purdue University, West Lafayette, Indiana 47907, USA}
\affiliation{School of Electrical and Computer Engineering, Purdue University, West Lafayette, Indiana 47907, USA}
\affiliation{Purdue Quantum Center, Purdue University, West Lafayette, Indiana 47907, USA}
\affiliation{Birck Nanotechnology Center, Purdue University, West Lafayette, Indiana 47907, USA}
    
\author{Zhang-qi Yin}\email{yinzhangqi@mail.tsinghua.edu.cn}
\affiliation{Center for Quantum Information, Institute for Interdisciplinary Information Sciences, Tsinghua University, Beijing 100084, China}


\date{\today}


\maketitle 


\section{Floquet formalism}\label{sec:floquet}

According to Floquet theorem, the solutions to the Sch\"ordinger equation with $T = 2\pi/\omega$-periodic Hamiltonian
\begin{equation}
	i\hbar\frac{\partial}{\partial t}|\psi(t)\rangle = H(t)|\psi(t)\rangle
\end{equation} can be written as superposition of the Floquet states 
\begin{equation}\label{eq:state}
	|\psi(t)\rangle = \sum_{\alpha}\langle\Psi_\alpha(t')|\psi(t')\rangle,
\end{equation}
for any $0\le t'<T$. The Floquet states are given by
\begin{equation}\label{eq:fstates}
	|\Psi_\alpha(t)\rangle = e^{-i\lambda_\alpha t}|\lambda_\alpha(t)\rangle,
\end{equation}
with real quasienergies $\lambda_\alpha$ and $T$-periodic part $|\lambda_\alpha(t)\rangle$. By substituting a Floquet solution $|\Psi_\alpha(t)\rangle$ into the Schr\"odinger equation, we find that the quasienergies and periodic part states follow the equation
\begin{equation}
	H_F(t)|\lambda_\alpha(t)\rangle = \lambda_\alpha |\lambda_\alpha(t)\rangle,
\end{equation}
where $H_F(t)$ is the Floquet Hamiltonian
\begin{equation}
	H_F(t) = H(t) -i\hbar\frac{\partial}{\partial t}.
\end{equation}
It can be shown that $\lambda_\alpha$ are uniquely determined only up to a multiple of $\hbar\omega$, i.e., $\lambda_\alpha + n\hbar\omega$ is also a quasienergy, corresponding to the periodic state $e^{-i n\omega t}|\lambda_\alpha(t)\rangle$. 

The quasi-energies are obtained by diagonalizing the Floquet matrix, which is the Floquet Hamiltonian $H_F(t)$ in the basis of Hilbert space $\mathcal{\Gamma}\otimes\mathcal{H}$. Here the $\mathcal{\Gamma} = \{e^{i n \omega t}|n=0,\pm1,\pm2,...\}$ is the Hilbert space of square integrable $T$-periodic functions, and the $\mathcal{H}$ is the original Hilbert space, i.e. the spin states in our model. The Floquet matrix is infinite dimensional, and we should truncate at some finite dimension, which is the case for our Hamiltonian with an external magnetic field Eq. (5) of the main text. If the Floquet matrix is block-diagonal, we can diagonalize each block matrix and get analytic solution, which is the case for our Hamiltonian Eq. (2) of the main text.

In the absence of an external magnetic field, the periodic Floquet states at $t = 0$ and quasienergies are given by the eigenvectors of the time-independent Hamiltonian in the interaction picture Eq. (2) of the main text and corresponding eigenvalues, denoted by $|\lambda_n\rangle$ and $\lambda_n$, $n = 0,\pm1$. According to Eq. \eqref{eq:fstates}, the Floquet states are given by the time evolution of $|\lambda_n\rangle$ multiplied by the phase factor of quasienergies, which reads
\begin{equation}\label{sup:eq:UF}
 e^{-i\lambda(t)}|\lambda_n (t)\rangle = e^{-i\lambda(t)}U(t)|\lambda_n\rangle = e^{-i\lambda(t)}e^{-i\lambda_n t}e^{i\omega S_z t}|\lambda_n\rangle.
\end{equation}
Substituting into Eq. \eqref{eq:state}, we get the solution of the Floquet Hamiltonian Eq. (1) of the main text.

\section{Spin dynamics and Effective Rabi Oscillation}
Here we present the numerical result for the solutions to the Schrödinger with Hamiltonian Eq. (1) of the main text, using the method discussed in the previous section. The spin is assumed to be initially prepared to the state $|0\rangle$. The time evolution of the levels population is plotted 

\begin{figure}
	\centering
	\includegraphics[width=.75\linewidth]{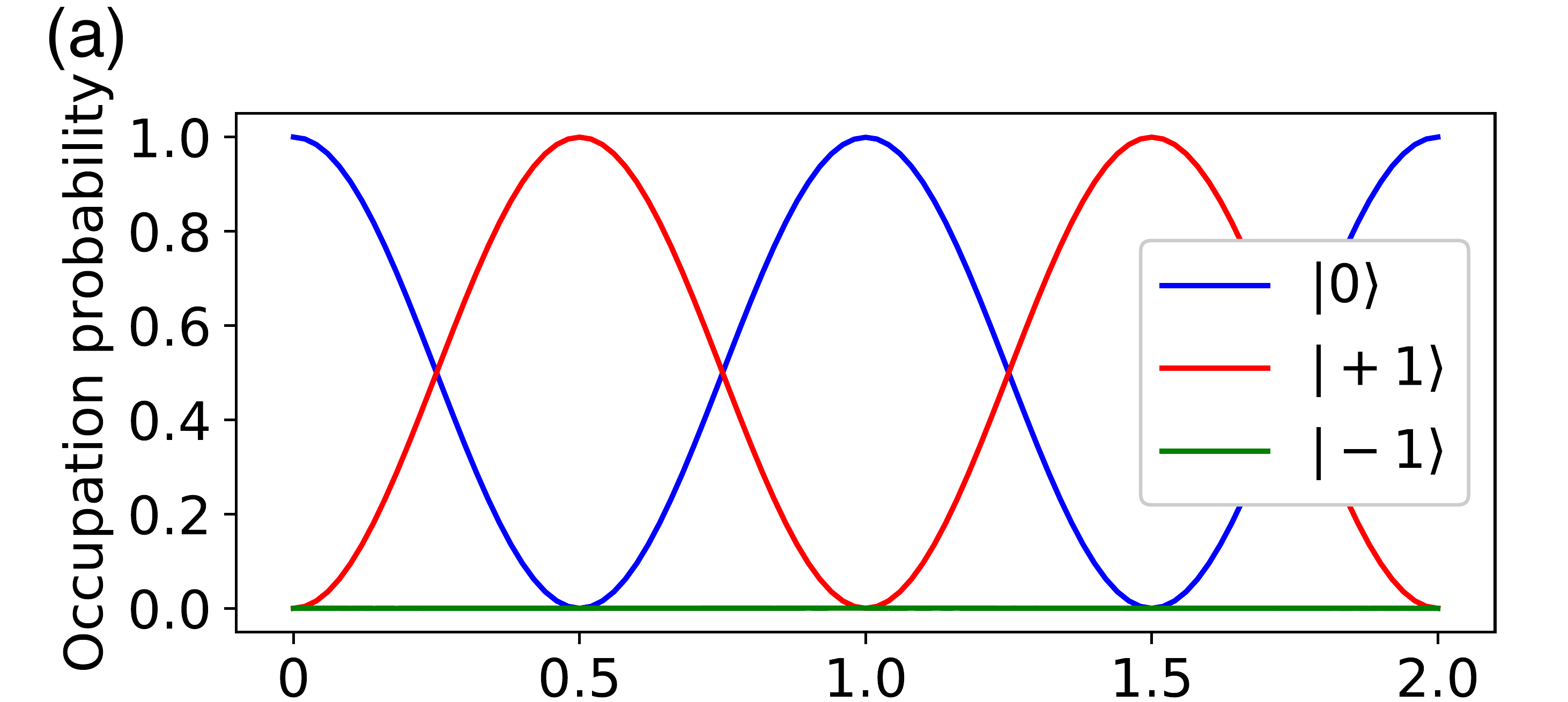}
	\includegraphics[width=.75\linewidth]{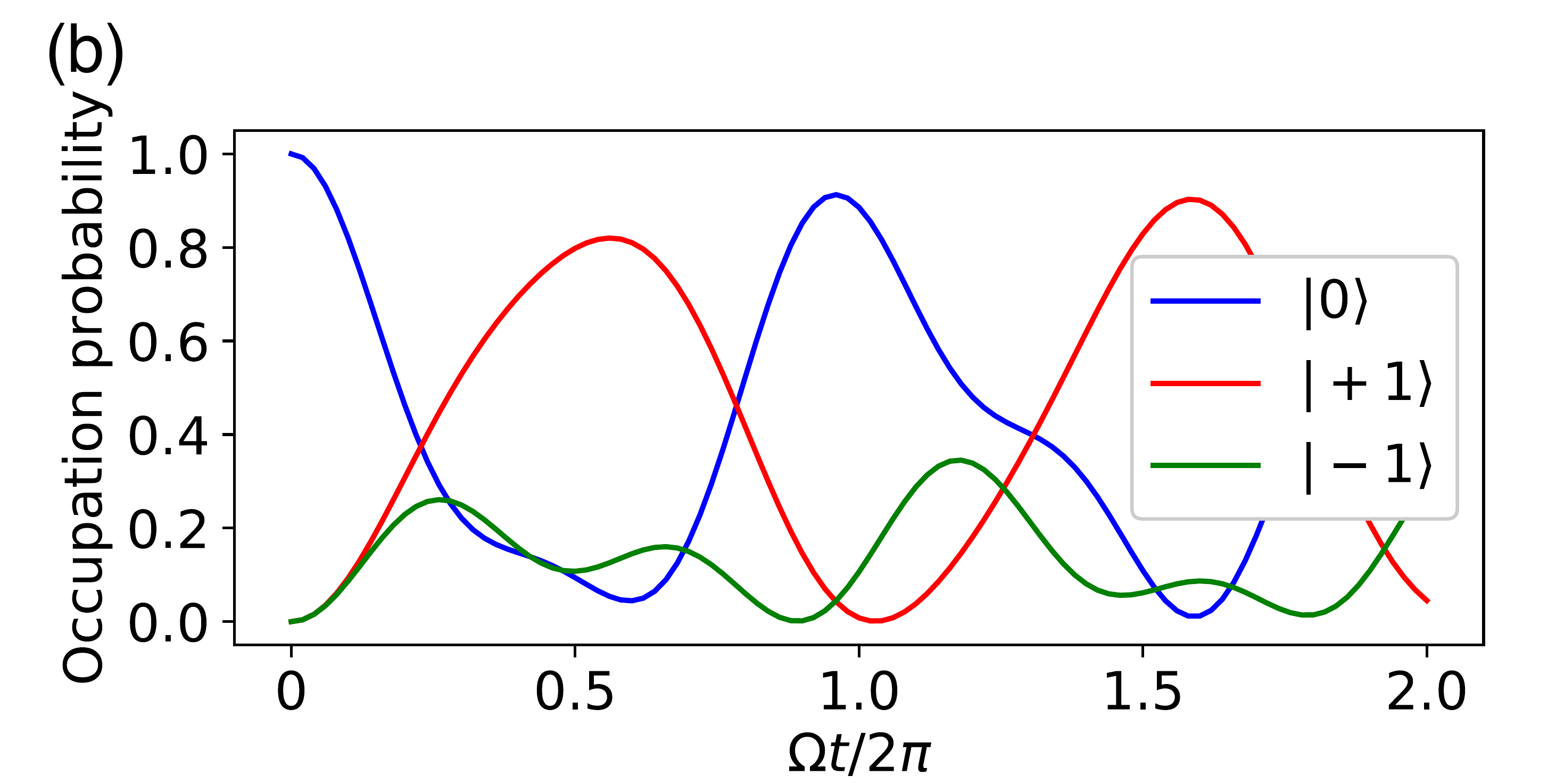}
	\caption{The time evolution of the spin levels population when the resonant condition $\omega=\omega_+$ fulfills. The spin is prepared to the state $|0\rangle$ initially.  The relative angle the solid spin to the angular velocity $\theta$ are (a) $\pi/100$ and (b) $\pi/4$.}
	\label{fig:noB}
\end{figure}

In the limit $\theta$ is much less than $1$, the electron spin dynamics can be described perfectly with effective Hamiltonian 
$$H_{\mathrm{eff}}=\begin{pmatrix}
0 & -\Omega/2 \\
-\Omega/2 & 0
\end{pmatrix}.$$
In Fig. \ref{fig:noB}(a), we take $\theta = \pi/100$, at the resonant frequency $\omega_+ = D/\cos\theta$. The numerical results show that the $|0\rangle$ and $|+1\rangle$ undergoes almost perfect Rabi oscillation, while the population in $|-1\rangle$ nearly does not change. The numerical results are consistent with the prediction of the effective Hamiltonian $H_{\mathrm{eff}}$.

In  Fig. \ref{fig:noB}(b), we plot a case for a large $\theta$, e.g. $\theta = \pi/4$, at the resonant frequency. It is found that the oscillation is slightly deviate from sinusoidal and the dynamics is not limited to the $|0\rangle$ and $|+1\rangle$ subspace.

From Eq. (2) in the main text, the $|\pm1\rangle$ states degenerate at $\theta = \pi/2$ and both couple with the $|0\rangle$. By adiabatically eliminating the level $|0\rangle$, the effective coupling between $|+1\rangle$ and $|-1\rangle$ causes a Rabi oscillation between them. The Rabi frequency can be obtained from Eq. (3) in the main text at $\theta = \pi/2$ with the perturbative method. The quasienergies difference corresponding to $|\pm1\rangle$, shown in Fig. (2)b in the main text, is given by the Rabi frequency $\Omega_{\pm1} = \omega^2/D$. This Rabi oscillation is much slower than $\omega$ at the adiabatic limit, but the spin mixing causes the geometric phase to jump between $|\pm1\rangle$. This explains the reason why Fig. (3)b in the main text has crossing near $\theta = \pi/2$.

\section{Hamiltonian of a rotating with external magnetic field}
If the NV center is rotating in the presence of an non-zero external static magnetic field, the total Hamiltonian reads \begin{equation}
H(t) = R(t)H_0 R^\dagger(t) + g \mu_B B\bm{S}\cdot\bm{n},
\end{equation}
where $g$ is the $g$-factor of the solid spin, $B$ is strength of magnetic field with direction $\bm{n}$.
To simplify the problem, we assume that the magnetic field is uniform and along the rotation $z$ axis.
In this way, the total Hamiltonian including magnetic field becomes
$H(t) = R(t) DS_z^2 R^\dagger(t) + g \mu_B B S_z$, 
We can imagine that the magnetic field is rotating with the solid spin and write the equivalent Hamiltonian as $H(t) = R(t)(DS_z^2 + g \mu_B B \bm{S}\cdot\bm{n})R^\dagger(t) 
= R(t)H_1 R^\dagger(t)$,
where $\bm{n} = (\sin\theta,\cos\theta,0)$ is the unitary vector along the magnetic field direction in the rotating frame, and $\Delta = -g\mu_B B$.
The time evolution of this Hamiltonian could be numerically solved by the Floquet formalism in section \ref{sec:floquet}.

\section{Non-adiabatic geometric phase for periodic Hamiltonian}
The Floquet theorem in section \ref{sec:floquet} implies that the time-evolution operator for a periodic Hamiltonian can be expressed as 
\begin{equation}\label{sup:eq:U}
	U(t) = Z(t)e^{i M t}
\end{equation}
where $Z(t)$ is a unitary $T$-periodic operator, i.e. $Z(T) = Z(0) = 1$, and $M$ is a Hermitian operator.
According to \cite{PhysRevLett.58.1593,Moore1990,Moore1990a}, the non-adiabatic geometric phase for the Floquet states (which are cyclic) are given by
\begin{equation}\label{sup:eq:gp}
	\gamma_n = i \int_0^T \langle \lambda_n | Z(t)^\dagger \frac{d}{dt} Z(t)|\lambda_n\rangle dt.
\end{equation}
For the zero-field Hamiltonian, the evolution operator can be expressed as $U(t) = Z(t)e^{iM t}$ Eq. \eqref{sup:eq:UF} where $Z(t) = e^{i\omega S_z t}$ is the unitary $T$-periodic operator, and $M = \tilde{H}_I$ is the Hermitian operator in Eq. \eqref{sup:eq:U}.

Since we have remove the geometric phase into the dynamical phase by the method in \cite{Giavarini1989}. If we want to recover the geometric phase from the dynamical phase, $Z(t)$ in Eq. \eqref{sup:eq:gp} should include an additional rotational transformation. As we have demonstrated in the main text using Fig. (1)b, $Z(t)$ should be chosen as the unitary transform $W^\dagger(t)$, which reads $Z(t) = W^\dagger(t)$. Then the geometric phases are given by Eq. (4) in the main text.

%